\documentclass[]{aa}
\usepackage{graphicx}
\usepackage[intlimits,sumlimits]{amsmath}
\usepackage{times,epsfig} 
\usepackage{natbib}
\usepackage{amssymb}
\bibpunct{(}{)}{;}{a}{}{,} 
\usepackage{amsmath}
\usepackage[english]{babel}
\begin{document}
\title{Reducing the gravitational lensing scatter of Type Ia supernovae without introducing any extra bias}
\titlerunning{Reducing the lensing scatter of SNe~Ia without introducing any extra bias}
\authorrunning{J\"onsson et al.}
\author{
J.~J\"onsson\inst{1}
\and E. M\"ortsell\inst{2}
\and J. Sollerman\inst{3,4}
}
\institute{University of Oxford Astrophysics, Denys Wilkinson
  Building, Keble Road, Oxford OX1 3RH, UK 
\and
Physics Department, Stockholm University, AlbaNova University Center, SE-10691 
Stockholm, Sweden 
\and
Stockholm Observatory, AlbaNova, Department of Astronomy, SE-10691 
Stockholm, Sweden 
\and
Dark Cosmology Centre, Niels Bohr
Institute, University of Copenhagen, Juliane Maries Vej 30, DK-2100 
Copenhagen \O, Denmark
\email{jacke@astro.ox.ac.uk}
} 
\offprints{J. J\"onsson}
\date{Received -- / Accepted --}
%
\abstract{}
{Magnification and de-magnification
  due to gravitational lensing will contribute to the 
  brightness scatter of Type Ia supernovae (SNe~Ia). The purpose of
  this paper is to investigate the possibility to decrease this
  scatter by correcting individual SNe~Ia 
  using observations of galaxies in the foreground, 
  without introducing any extra bias.
}
{We simulate a large number of SN~Ia lines of sight populated by galaxies. 
 For each line of sight the true magnification factor and
 an estimate thereof are calculated.
 The estimated magnification factor corresponds to what an
  observer would infer from a survey like SNLS. Using the simulated
  data we investigate the possibility to estimate the magnification of
  individual supernovae with enough precision to be able to 
  correct their brightness for gravitational lensing with negligible bias.
}
{
Our simulations show that the bias arising from gravitational lensing
corrections of individual SNe~Ia is negligible for current and next 
generation surveys and that the scatter from lensing can be 
reduced by approximately a factor $2$. 
The total scatter in the SN~Ia magnitudes could be reduced by $4\%$ for an intrinsic dispersion
of 0.13 mag. For an intrinsic dispersion of $0.09$ mag, 
which might not be to unrealistic for future surveys, the total scatter could be reduced by $6\%$.
This will reduce the errors on cosmological parameters derived from supernova data 
by 4--8\%. The prospect of correcting for lensing is thus very good. 
}
{}

\keywords{supernovae: general -- gravitational lensing} %
\maketitle

\section{Introduction}
Large dedicated supernova surveys, such as
SNLS \citep{ast06}, ESSENCE \citep{mik07,woo07}, 
and SDSS-II \citep{fri07}, have recently gathered data on a large number 
of supernovae at cosmological redshifts.
Consequently, supernova
cosmology has now reached a stage where systematic uncertainties,
rather than lack of statistics, limit the ability to constrain 
cosmological models.
Extinction by dust and peculiar motions of host galaxies seem to
be the systematic effects which currently pose the largest
difficulties \citep[e.g.][]{lei08}. In this paper we investigate the
possibility to correct for  
another systematic uncertainty -- gravitational lensing. 

In an inhomogeneous universe like ours, light from
a distant Type Ia supernova (hereafter SN~Ia) is inevitably
affected by gravitational lensing. Weak gravitational lensing,
which is what we will consider in this paper, refers to
the phenomena which occur when a bundle of light rays is
distorted by the gravitational fields exerted by matter. The distortion 
can lead to magnification or de-magnification of the SN~Ia flux.
This (de)magnification 
can be described by the magnification factor, $\mu$, which depends on
the distribution and composition of the foreground matter.
Both baryonic and dark matter influence the trajectories of photons, but  
since the latter dominates on the scales relevant for this work, 
we only use the baryonic
component as a tracer of the dark matter.
If an observer residing in a homogeneous universe would measure the flux $f$, 
an observer situated in an inhomogeneous universe will measure the flux $f_{\rm obs}=\mu f$.
Due to flux conservation, as long as we do not have multiple images,
 the probability distribution function for the magnification
 factor, $P(\mu)$,
satisfies the constraint $\langle \mu \rangle =\int P(\mu)\mu d\mu=1$, 
i.e.~the average magnification factor is unity. 
The effect of gravitational lensing consequently average out, which
implies that 
this systematic effect can be controlled by observing large numbers of SNe~Ia.
In reality however, $\langle \mu \rangle$ converges very slowly towards unity as the 
sample size increases \citep{hol05}. This owes to the fact that the whole distribution 
of $P(\mu)$, including its elusive high magnification tail, must be sampled,
which in addition to large numbers also requires large survey areas \citep{cor06}.

Although the bias in principle can be beaten down by large numbers,
gravitational lensing still contributes to the 
brightness scatter of SNe~Ia.
This contribution increases with 
redshift and is expected to be substantial at $z \ga 1$ \citep{hol05}.  
For precision measurements of cosmological parameters it is therefore desirable
to minimize this scatter by measuring and correcting for the
gravitational lensing effect.

The method we rely upon is based on modeling the dark matter haloes, which 
surrounds  galaxies, using the observed properties of the galaxies. This method
takes into account contributions to the lensing from sub-arc-minute scales, which makes
it a viable method in contrast to shear maps \citep{dal03} which are insensitive to these scales. 
The possibility to correct individual SNe~Ia for gravitational lensing
using the method considered here
was investigated by \cite{gun06}, where simulations showed that the
lensing scatter could be reduced by almost a factor $3$ for a source at
redshift $z=1.5$. In this paper we focus on the possibility to correct
for gravitational lensing at the lower redshifts ($z \la 1$)
accessible to the aforementioned large ground based surveys. 
The simulations aim at mimicking the supernova legacy survey \cite[SNLS,][]{ast06}, and follow
closely the strategy by \citet[][hereafter J08]{jon08}.

For a correction to be
useful the following two criteria must be met:
\begin{enumerate}
\item The correction should decrease the scatter in SN~Ia magnitudes due to gravitational
  lensing.
\item Any bias introduced in the SN~Ia magnitudes from the correction should be negligible.
\end{enumerate}
In Sect.~\ref{sec:sim}, we use simulated data to 
investigate the possibility to perform corrections satisfying the
two criteria above. 
The method we use to correct for lensing
is explained and applied to the simulated data in Sect.~\ref{sec:corr}. 
Our results are discussed and summarized in 
Sect.~\ref{sec:disc}.

\section{Simulations \label{sec:sim}}
Since our simulations are very similar to the ones presented in
more detail by J08, we only provide a brief presentation here.

We have simulated deep galaxy catalogs with properties 
(galaxy positions, redshifts, luminosities, types)
based on real observations \citep{dah05}. For a single line of sight we first 
compute the true magnification factor, $\mu_{\rm true}$, 
using the multiple lens plane algorithm implemented in the Q-LET package \citep{gun04}. All galaxies are 
included in the calculation and the dark matter haloes surrounding each 
galaxy are modeled by truncated Navarro-Frenk-White \citep[NFW,][]{nav97} profiles.
Then we compute the magnification factor which an observer 
(with access to SNLS data) would estimate, $\mu_{\rm est}$, 
for the same line of sight. The observer would, for example, only see galaxies brighter
than the magnitude limit ($i'_{AB} \le 25.5$) of the survey and would probably have access only to
photometric redshifts (with a precision of $\sigma_{\Delta z/(1+z)} \simeq 0.03$ 
for most galaxies, but failing catastrophically for 2--6\% of the galaxies).
The content of the line of sight is therefore rather different
when the true and estimated magnifications are calculated. 
Since the observer does not know what the correct model is,
there is also a difference in the model used to compute the true and the estimated 
magnification factor.  
Since the effects of gravitational lensing are redshift dependent, we
simulate gravitational lensing of sources in the redshift range $0.2
\le z \le 1.1$, relevant for e.g.~the SNLS survey.

The result of the simulations are hence pairs of true, $\mu_{\rm true}$, and
estimated, $\mu_{\rm est}$, magnification factors.
Since most cosmology fits to date have been performed
using magnitudes rather than fluxes, we treat magnification in terms
of logarithmic units. Throughout the paper we denote true and
estimated magnifications by $\Delta m_{\rm true}=-2.5\log_{10}\mu_{\rm true}$ 
and $\Delta m_{\rm est}=-2.5\log_{10}\mu_{\rm est}$, respectively.

\begin{figure} 
\resizebox{\hsize}{!}{\includegraphics[angle=-90]{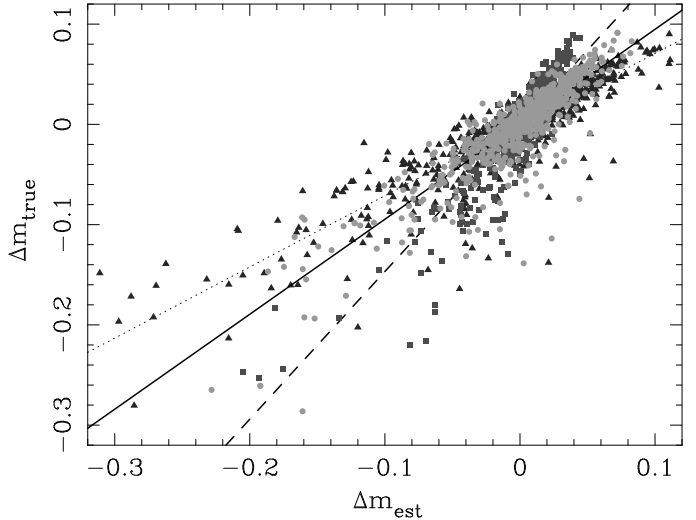}}
\caption{\label{fig:scatter} True magnification, $\Delta m_{\rm true}$, 
  versus estimated magnification, $\Delta m_{\rm est}$,
  for three different scenarios: no shift in halo masses (circles), 
  underestimated halo masses (squares), and 
  overestimated halo masses (triangles).
  The straight lines show the best fits
  to the no shift (solid line), underestimation (dashed line), and 
  overestimation (dotted line) scenario.
}
\end{figure}

Figure~\ref{fig:scatter} shows
simulated pairs of magnification factors in logarithmic units.
This plot shows most of the simulated pairs, but not the 
small fraction ($\lesssim 0.3\%$)  belonging
to the high magnification tail with $\Delta m_{\rm true} \lesssim -0.3$.
The scatter in this plot comes mostly from 
the uncertainty in the model used to 
convert galaxy luminosity to mass (in our case Tully-Fisher and Faber-Jackson relations). 
Given the importance of the luminosity-to-mass
model, we continue this investigation by considering
three different scenarios, all shown in Fig.~\ref{fig:scatter}:
\begin{enumerate}
\item No shift in halo masses (circles).
\item Underestimated halo masses (squares).
\item Overestimated halo masses (triangles).
\end{enumerate}
If the masses of the dark matter haloes are substantially overestimated 
(by 50$\%$ in scenario 2) or underestimated (by 50$\%$ in scenario 3), 
the gravitational magnification estimates will be erroneous.
Our simulations aim to investigate to what 
extent a proper correction can be made for the lensing magnification given these uncertainties.
Other sources of uncertainty, such as scatter in 
the luminosity-to-mass model, were also included for all three scenarios. 
Clearly, the distributions
of points in the plot differ for the three scenarios.
True and estimated magnification factors are apparently correlated in
all scenarios, albeit with different slopes. The no shift scenario has a slope 
(outlined by the solid line in Fig.~\ref{fig:scatter})
near unity, which
reflects the fact that the model used to estimate the magnification is
very close to the correct one. The underestimation (dashed line) and
overestimation (dotted line) scenarios have slopes which differs from unity. 

From Fig.~\ref{fig:scatter}, the asymmetry of both the distribution of 
$\Delta m_{\rm true}$ and $\Delta m_{\rm est}$ is evident. 
For slopes differing from unity this asymmetry implies an asymmetry in
the errors in the estimated magnifications.
The estimated magnification of a SN~Ia in the high
magnification tail (negative values in Fig.~\ref{fig:scatter}) is likely to 
be more erroneous than a typical SN~Ia which is slightly de-magnified 
(positive values in Fig.~\ref{fig:scatter}).

%

\section{Gravitational lensing corrections \label{sec:corr}}
\subsection{The magnification-residual-diagram}
Since gravitationally magnified (de-magnified) SNe~Ia should be
brighter (fainter) than the average supernova, a correlation between
estimated magnification and SN~Ia brightness is expected.
According to Fig.~\ref{fig:scatter}, which can be seen as an idealized 
magnification-residual-diagram, a correlation 
is expected even if the model used to estimate the magnification
factors is biased. 
A real magnification-residual-diagram, 
where  Hubble diagram residuals are used instead of $\Delta m_{\rm true}$,
will also be smeared by, e.g., measurement errors and 
intrinsic brightness scatter. 
We take the supernova Hubble diagram residuals 
to be the differences between observed magnitudes, $m_{\rm obs}$, and
magnitudes predicted by a cosmological model 
of a homogeneous universe (we assume a flat universe, dominated 
by a cosmological constant, with $\Omega_{\rm M}=0.3$ and 
$H_0=70$ km~s$^{-1}$~Mpc$^{-1}$).  
In the following, we assume that the residuals have been
computed using the correct cosmological model. The residual is then
the sum of $\Delta m_{\rm true}$ and noise, $\Delta m_{\rm noise}$, from,
e.g.,~intrinsic SN~Ia 
brightness scatter and measurement errors. We ignore other 
systematic sources of uncertainty and assume the noise to be Gaussian.

\subsection{Magnitude bias}
\begin{figure}
\resizebox{\hsize}{!}{\includegraphics[angle=-90]{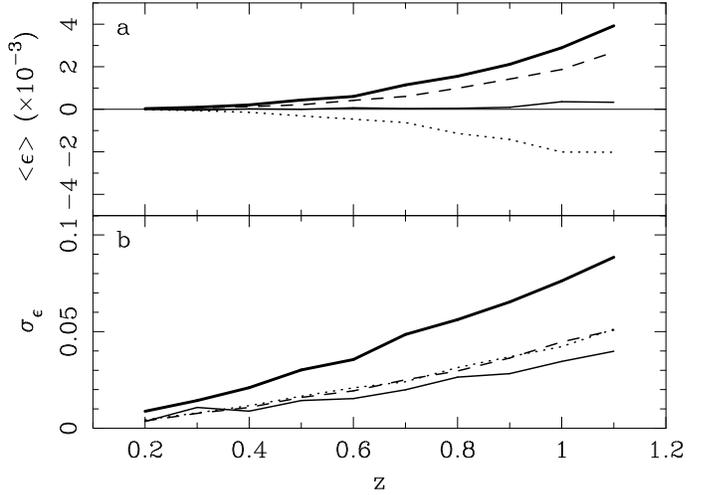}}
\caption{\label{fig:corrplot} Bias (panel a) and dispersion (panel b)
  in $\epsilon$ as a function of redshift.
  The no shift scenario is represented by
  the thin solid curves. Dashed and dotted curves correspond
  to the under- and overestimation scenarios, respectively.
  Bias and dispersion of $\Delta m_{\rm true}$,
  outlined by the thick solid curves, are also shown for comparison. The
  estimated magnifications, $\Delta m_{\rm est}$, were used to correct for gravitational
  lensing.
}
\end{figure}

Because of gravitational lensing, an observer in an inhomogeneous universe 
measures the magnitude
\begin{equation}
m_{\rm obs}=m+\Delta m_{\rm true},
\end{equation}
rather than the magnitude $m=-2.5\log_{10}f$, which would 
be measured if the observer resided in a homogeneous universe. 
The effect of gravitational lensing average out because the average
magnification factor is unity.  
However, an infinite number of SNe~Ia is in principle 
required, since 
the average magnification factor 
of a finite sample converges very slowly to unity as the sample size increases
 [see Fig.~2 in \citet{hol05}]. 
When magnitudes are used instead of fluxes, on the other hand, the effect no
longer average out even in principle, because $\langle \log_{10} \mu \rangle \ne 0$
even though $\langle \mu \rangle=1$. Observed magnitudes are consequently
biased with an amount depending on the redshift.
The thick solid curve in 
Fig.~\ref{fig:corrplot}a shows this bias,
$\langle m_{\rm obs} \rangle -m=\langle \Delta m_{\rm true} \rangle$, which 
is of the order $10^{-3}$.

Recently \citet{sar07} investigated the expected bias 
in the dark energy equation of state parameter
due to gravitational lensing for
future SNe~Ia surveys. In their study \citet{sar07} considered both
averaging over magnitudes (as we do here) and a flux-averaging 
technique \citep{wan00,wan04}. In both cases they found the resulting 
bias in the equation of state parameter to be negligible, compared to the
expected statistical uncertainty. 

%

\subsection{Lensing corrections \label{sec:corr2}}
If the estimated magnification factors are used to correct
for gravitational lensing, the corrected magnitudes would be
\begin{equation}
m_{\rm corr}=m_{\rm obs}-\Delta m_{\rm est}=m+\epsilon,
\label{eq:corr}
\end{equation} 
where the correction is characterized by the difference 
\begin{equation}
\epsilon=\Delta m_{\rm true}-\Delta m_{\rm est}.
\label{eq:eps}
\end{equation}
For the correction to be useful, according to the first criterion stated
above, the dispersion in the SN~Ia magnitudes must be decreased 
($\sigma_{m_{\rm corr}} < \sigma_{m_{\rm obs}}$), which 
consequently means that 
the dispersion in $\epsilon$ must be smaller than the dispersion in 
$\Delta m_{\rm true}$, i.e.~ $\sigma_{\epsilon}/\sigma_{\Delta m_{\rm true}} <1$.

The second criterion requires that  the correction does not introduce 
any extra bias in the corrected magnitudes. According to
Eq.~(\ref{eq:corr}) this bias is given by $\langle \epsilon \rangle$. 
Whether the bias is negligible or not depends on the 
size of the statistical error. We require 
$\langle \epsilon \rangle /\sigma_{\Delta m_{\rm true}}<1$ 
for a correction to be considered useful.

Figures~\ref{fig:corrplot}a~and~\ref{fig:corrplot}b show the bias and 
dispersion, respectively,
for $\epsilon$ computed for the three different scenarios 
discussed in Sect.~\ref{sec:sim} over a range 
of redshifts. In this case, no noise has been taken into account.
The scenario corresponding to no shift in halo masses is shown by the thin solid curves.
Thin dashed and dotted curves outline the results for 
the scenarios were halo masses have been under- and overestimated, 
respectively.
We also display with the thick solid curves the bias and dispersion in 
$\Delta m_{\rm true}$. 
For all three scenarios, the first criterion above is fulfilled, 
i.e.~$\sigma_{\epsilon} < \sigma_{\Delta m_{\rm true}}$ at all redshifts.
At $z \simeq 0.8$, which is the redshift where the SNLS distribution of SNe~Ia peaks,
the reduction in the scatter is a factor  $2.1$, $1.9$, and $1.8$ for scenario 1, 2, and 3.
The absolute value of the bias, $|\langle \epsilon \rangle|$, is $\lesssim 3\times10^{-3}$ for all scenarios. If this
bias is divided by $\sigma_{\Delta m_{\rm true}}$, we find 
$|\langle \epsilon \rangle /\sigma_{\Delta m_{\rm true}}| \lesssim 4\times10^{-2}$, which 
implies successful corrections for all cases and at all redshifts 
according to our second criterion.
Moreover, for all three scenarios 
$|\langle \epsilon \rangle| < |\langle \Delta m_{\rm true} \rangle|$. 
We have thereby demonstrated that such a correction for
gravitational lensing would clearly decrease the scatter in the
Hubble diagram without adding significant bias after the corrections. 

Let us now investigate the potential 
benefits of gravitational lensing corrections in
the presence of realistic intrinsic brightness scatter and measurement errors.
We have simulated SNLS like data sets, consisting of $250$ SNe~Ia each,
for the three scenarios. 
Since the results depend on the assumed errors, we have also varied the
intrinsic brightness scatter in the simulations. To model the measurement errors
we have used the following fit to the first year SNLS data \citep{ast06}:
\begin{equation}
\sigma_{\rm err}=\left\{
\begin{array}{ll}
 0.05 \mbox{ mag} &\mbox{ if $z<0.8$} \\
 0.84z^2-1.04z+0.34 \mbox{ mag}  &\mbox{ if $z \geq 0.8$}.
       \end{array}
\right.
\label{eq:err}
\end{equation}
For each simulated data set,
$\sigma_{\Delta m_{\rm true}}$,
$\sigma_{\epsilon}$, and $\langle \epsilon \rangle$
 were computed.
Figure~\ref{fig:criteria} shows scatter plots of 
$\langle \epsilon \rangle/\sigma_{\Delta m_{\rm true}}$ (the second criterion)
versus
$\sigma_{\epsilon}/\sigma_{\Delta m_{\rm true}}$ (the first criterion)
for different assumed errors.
As in Fig.~\ref{fig:scatter} the three scenarios are indicated by different plotting symbols.
In Fig.~\ref{fig:criteria}a measurement errors, modeled by Eq.~({\ref{eq:err}), and intrinsic dispersion,
 $\sigma_{\rm int}=0.13$ mag, correspond to the precision of first year SNLS data. A future decrease
 in the intrinsic dispersion is not to unrealistic. We therefore show in Fig.~\ref{fig:criteria}b results obtained
 for the same measurement errors, but with $\sigma_{\rm int}=0.09$ mag. Figure~\ref{fig:criteria}c shows 
 results for $\sigma_{\rm int}=0.09$ mag and negligible measurement errors, i.e.~$\sigma_{\rm err}=0$ mag. 
 For comparison we also show results in Fig.~\ref{fig:criteria}d for an intrinsic dispersion of only 
 0.05 mag and $\sigma_{\rm err}=0$.
The distributions of points 
in the scatter plots are rather similar irrespective of the scenario.
A value of $\sigma_{\epsilon}/\sigma_{\Delta m_{\rm true}}$ 
less than unity, i.e.~to the left  of the vertical lines, 
indicates a successful correction according to the first criterion. 
The number of simulated data sets failing to meet the first criterion,
i.e.~points to the right of the vertical line, decreases as 
the errors decrease. In Fig.~\ref{fig:criteria}a the
correction fails for 4\%, 1\%, and 10\% of the synthetic data sets
for the no shift, underestimation, and overestimation scenario. These numbers
drop to less than one percent for Fig.~\ref{fig:criteria}d.

From Fig.~\ref{fig:criteria} it is also evident that the bias after the correction
has been performed is small; for all cases 
$|\langle \epsilon \rangle/\sigma_{\Delta m_{\rm true}}| < 0.25$.  
According to Fig.~\ref{fig:criteria}, corrections for gravitational lensing are likely
to reduce the scatter in SN~Ia brightness without introducing any harmful bias.

Figure~\ref{fig:std} shows a projection of Fig.~\ref{fig:criteria}
focusing on the distribution of the ratio
$\sigma_{\epsilon}/\sigma_{\Delta m_{\rm true}}$.
Figures~\ref{fig:std}a,~\ref{fig:std}b, and~\ref{fig:std}c show results
obtained for the no shift, under-, and overestimation scenarios.  
Solid and dashed curves correspond to measurement errors modeled
by Eq.~(\ref{eq:err}) and intrinsic dispersion of $0.13$ and $0.09$ mag, respectively.
Dotted and dash-dotted curves correspond to negligible measurement errors
($\sigma_{\rm err}=0$) and  intrinsic dispersion $0.09$ and $0.05$ mag, respectively.
In Table~\ref{tab:std}, the characteristics of the distributions
shown in Fig.~\ref{fig:std} are summarized.
From Fig.~\ref{fig:std},
it is clear that the distributions obtained
with systematic shifts in the halo masses (scenario 2 and 3) are rather similar to the 
distributions corresponding to the no shift scenario. 
From Table~\ref{tab:std} we see that 
$0.96 \leq \langle \sigma_{\epsilon}/\sigma_{\Delta m_{\rm true}} \rangle \leq0.97$
when the intrinsic scatter is $\sigma_{\rm int}=0.13$ mag, 
which implies that the average effect of correcting for gravitational
lensing is to decrease the total scatter by 3--4\%. 
A realistic future intrinsic dispersion of $\sigma_{\rm int}=0.09$ mag would allow 
a decrease in the total scatter by 5--6\%, for measurement errors modeled by 
Eq.~(\ref{eq:err}), and by 7--8\% if measurement errors could be reduced to a negligible level.

We have also investigated the possible improvements on cosmological 
parameter estimation which could be gained from
gravitational lensing corrections. 
Using the Fisher information matrix, we computed confidence ellipses in the 
$(\Omega_{\rm M},w)$-plane assuming a flat universe. The parameters 
$\Omega_{\rm M}$ and $w$ denote the dimensionless matter density of the universe and 
a constant dark energy equation of state parameter, respectively.  In this analysis,
a low and a high redshift SN~Ia data set were used. The high redshift data set 
consists of 250 SNe~Ia with redshifts and measurements errors similar to
first year SNLS data \citep{ast06}. The purpose of the low redshift data set, consisting of 
44 SNe~Ia unaffected by lensing, is to anchor the Hubble diagram.
Redshifts and uncertainties  were similar to the low redshift data set used in \citet{ast06}.
Errors due to gravitational lensing before and after correction were modeled using 
the curves in Fig.~\ref{fig:corrplot}. Our treatment of the errors is hence simplified, because these curves 
correspond to the dispersion and does not take the asymmetry of the distributions into account.
Table~\ref{tab:fisher} shows the
relative improvement in the  area of the confidence level ellipses due to gravitational 
lensing corrections, $A_{\rm corr}/A_{\rm lensed}$. 
The scatter due to gravitational lensing is rather small compared to intrinsic brightness dispersion and 
measurement errors. Nevertheless, corrections for gravitational lensing would result in 
improvements of 4--8\% for realistic values of the intrinsic dispersion and $\sim 30\%$ for 
a very optimistic scenario where other sources of error are negligible and the supernovae are 
calibrated to $5\%$ accuracy.

%
\begin{figure}
\resizebox{\hsize}{!}{\includegraphics[angle=-90]{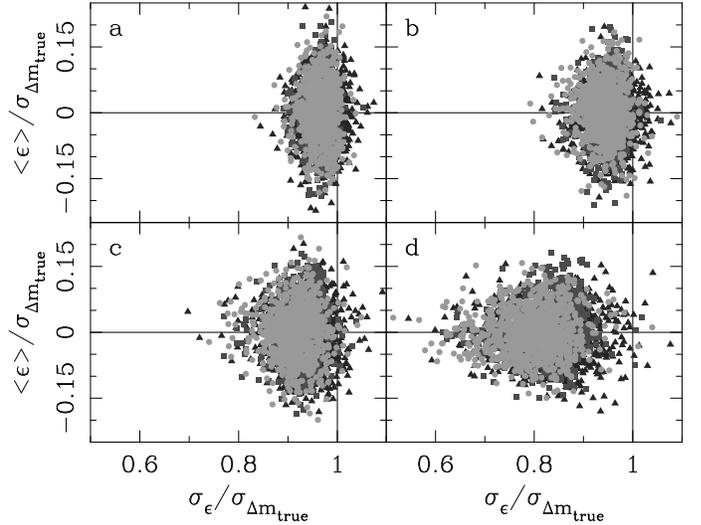}}
\caption{\label{fig:criteria} Scatter plot of 
$\langle \epsilon \rangle/\sigma_{\Delta m_{\rm true}}$ versus
$\sigma_{\epsilon}/\sigma_{\Delta m_{\rm true}}$ 
 for simulated SNLS like data sets
consisting of 250 SNe~Ia each. 
Circles, squares, and triangles  represent the no shift, under-, and overestimation 
scenarios, respectively.
The different panels correspond to different
intrinsic brightness dispersion and measurement errors: 
 Panel a) $\sigma_{\rm int}=0.13$ mag, $\sigma_{\rm err}$ given by Eq.~(\ref{eq:err});
Panel b) $\sigma_{\rm int}=0.09$ mag, $\sigma_{\rm err}$ given by Eq.~(\ref{eq:err});
Panel c) $\sigma_{\rm int}=0.09$ mag, $\sigma_{\rm err} = 0$ mag;
Panel d) $\sigma_{\rm int}=0.05$ mag, $\sigma_{\rm err} = 0$ mag.
}	
\end{figure}

%
\begin{figure}
\resizebox{\hsize}{!}{\includegraphics[angle=-90]{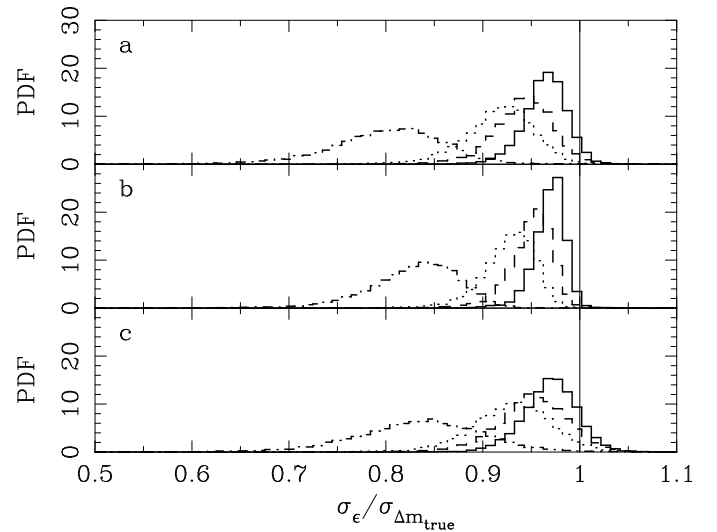}}
\caption{\label{fig:std} Probability distribution functions of 
  $\sigma_{\epsilon}/\sigma_{\Delta m_{\rm true}}$ for different scenarios 
  and errors. The distributions
  were obtained for
  simulated SNLS like data sets consisting of $250$ SNe~Ia. 
  Panel a, b, and c correspond to the no shift, under-, and overestimation scenario.
  The curves show results for different assumed errors:
  Solid curves correspond to $\sigma_{\rm err}$ given by Eq.~(\ref{eq:err}) and $\sigma_{\rm int}=0.13$ mag; 
  Dashed curves correspond to $\sigma_{\rm err}$ given by Eq.~(\ref{eq:err}) and $\sigma_{\rm int}=0.09$ mag; 
  Dotted curves correspond to $\sigma_{\rm err}=0$ and $\sigma_{\rm int}=0.09$ mag; 
  Dash-dotted curves correspond to $\sigma_{\rm err}=0$ and $\sigma_{\rm int}=0.05$ mag.
}	
\end{figure}

%
\begin{table}
\begin{minipage}[t]{\columnwidth}
\caption{Characteristics (mean and root mean square) of the distribution of 
$\sigma_{\epsilon}/\sigma_{\Delta m_{\rm true}}$ for the no shift, under- and 
overestimation scenarios and different assumed errors. The distributions were obtained from $10000$
  simulations of SNLS like data sets consisting of 250 SNe~Ia each.}         
\label{tab:std}      
\centering                                      
\renewcommand{\footnoterule}{}  
\begin{tabular}{llllllll}          
\hline\hline                        
$\sigma_{\rm int}$ & $\sigma_{\rm err}$  &
\multicolumn{2}{c}{Scenario 1\footnote{No shift in halo masses.}}& 
\multicolumn{2}{c}{Scenario 2\footnote{All halo masses are underestimated
    by 50\%.}}& 
\multicolumn{2}{c}{Scenario 3\footnote{All halo masses are overestimated
   by 50\%.}}  \\
(mag) & (mag) & Mean & RMS & Mean & RMS & Mean & RMS \\
\hline
$0.13$ & Eq.~(\ref{eq:err}) & 0.96 & 0.02 & 0.97 & 0.02 & 0.97 & 0.03
\\
$0.09$ & Eq.~(\ref{eq:err}) & 0.94 & 0.03 & 0.95 & 0.02 & 0.95 & 0.04
\\
$0.09$ & 0 & 0.92 & 0.04 & 0.93 & 0.03 & 0.93 & 0.04 \\
$0.05$ & 0 & 0.80 &0.06 & 0.83 & 0.05 & 0.83 & 0.07 \\
\hline                                             
\end{tabular}
\end{minipage}
\end{table}

%
\begin{table}
\begin{minipage}[t]{\columnwidth}
\caption{Relative improvement in confidence level contours in the $(\Omega_{\rm M},w)$-plane from corrections
for gravitational lensing. The results presented in this table were obtained using a Fisher matrix analysis for
a SNLS like data set consisting of 250 high redshift SNe~Ia and 44 low redshift SNe~Ia (assumed not to be affected by lensing).}
\label{tab:fisher}      
\centering                                      
\renewcommand{\footnoterule}{}  
\begin{tabular}{llllll}          
\hline\hline                        
$\sigma_{\rm int}$ & $\sigma_{\rm err}$\  &
Scenario 1\footnote{No shift in halo masses.}& 
Scenario 2\footnote{All halo masses are underestimated
    by 50\%.}& 
    Scenario 3\footnote{All halo masses are overestimated
   by 50\%.}  \\
(mag) & (mag)& $A_{\rm corr}/A_{\rm lensed}$ & $A_{\rm corr}/A_{\rm lensed}$ & $A_{\rm corr}/A_{\rm lensed}$\\
\hline
 0.13  &  Eq.~(\ref{eq:err}) &    0.95   &      0.95    &       0.96 \\
  0.09   & Eq.~(\ref{eq:err}) &     0.92   &      0.93  &         0.93 \\
    0.09   & 0 &     0.84   &      0.87  &         0.87 \\
        0.05  & 0 &     0.69   &      0.74  &         0.74 \\
\hline                                             
\end{tabular}
\end{minipage}
\end{table}

\subsection{Correction coefficient}
In the next step, we could try to improve the correction
by considering a model which incorporates the
different slopes exhibited in Fig.~\ref{fig:scatter}. 
Since the points in Fig.~\ref{fig:scatter} appears to cluster around
straight lines with different slopes, a  linear correction 
in $\Delta m_{\rm est}$,
\begin{equation}
\Delta m_{\rm est}^{B}=B\Delta m_{\rm est},
\label{eq:corrlog}
\end{equation}
is the simplest approach. The $B$-coefficient is related to how 
bad we are at estimating the luminosity-to-mass relation, and correcting with 
$B$ is the most straightforward way to remedy this. In J08 it was shown that 
since $B$ is sensitive to the normalization of halo masses, we can constrain
galaxy masses using the fitted value of $B$. 

Since the effects of gravitational lensing are redshift dependent, the
correction coefficient could change with redshift.
Figure~\ref{fig:abplot} shows the best fit values of $B$ to a 
large number of simulated pairs
of $\Delta m_{\rm est}$ and $\Delta m_{\rm true}$ for our three 
scenarios as a function of redshift. 
An optimal correction would,
of course, use values of 
$B$ obtained at different redshifts,
but would be difficult to derive for a limited data set. Fortunately, the
correction coefficient $B$ 
stays rather constant with redshift, which implies that  a slope
fitted to a data set spanning a range of redshifts should be
useful.

Figure~\ref{fig:corrplot2}, which is
is analogous to Fig.~\ref{fig:corrplot}, shows dispersion and bias 
as a function of redshift after corrections 
have been performed using the best fitting correction coefficient, $B$, to the entire
data set, which spans a range of redshifts. Equation~(\ref{eq:corrlog})
can not remove the bias at all redshifts, but can bring the bias 
for the under- and overestimation scenario down by $\sim 25\%$ and $\sim 70\%$, respectively. 

Including the correction coefficient in the corrections have a negligible impact on $\langle \epsilon \rangle/ \sigma_{\Delta m_{\rm true}}$
for our simulated data sets (see Sect.~\ref{sec:corr2}), but 
make a difference to the distributions of 
$\sigma_{\epsilon}/\sigma_{\Delta m_{\rm est}}$. Figure~\ref{fig:std2}, which is similar to Fig.~\ref{fig:std}, shows the 
distributions for the no shift scenario when corrections have been performed using Eq.~(\ref{eq:corrlog}).
Only the distributions corresponding to scenario 1 is shown, because the distributions belonging to the three
different scenarios are indistinguishable. Equation~(\ref{eq:corrlog}) hence eliminates the
effect of the biased halo masses. 
Furthermore, all distributions are pushed  to the left of the vertical line 
($\sigma_{\epsilon}/\sigma_{\Delta m_{\rm est}}=1)$. The 
inclusion of the correction coefficient $B$, thus
reduces 
the corrected gravitational lensing dispersion down to the level of 
scenario 1, even if the halo masses are originally over- or underestimated. 

It also improves cosmological constraints in the $(\Omega_{\rm M},w)$-plane 
to the
same level as for the no shift scenario without correction coefficient (see Table~\ref{tab:fisher}). 
No further improvement in the cosmological constraints can be achieved for the no shift scenario by
Eq.~(\ref{eq:corrlog}). 

%
\begin{figure}
\resizebox{\hsize}{!}{\includegraphics[angle=-90]{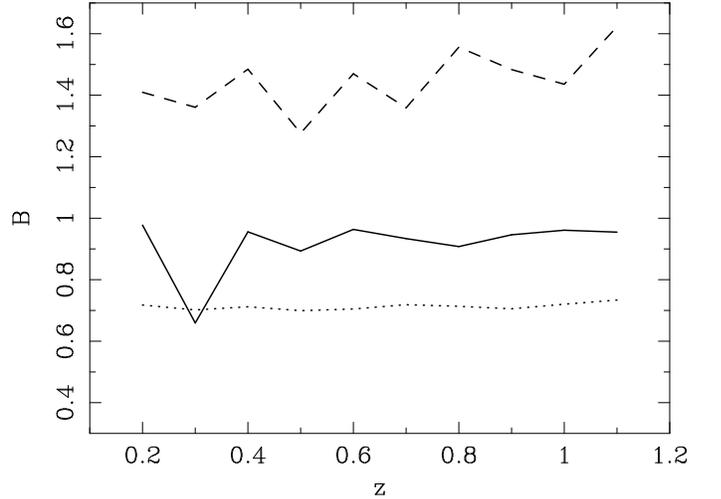}}
\caption{\label{fig:abplot}  Correction coefficient $B$ as a
  function of redshift for three simulated scenarios. 
   For the first scenario (solid curve) there is no systematic shift
  in halo masses. For the second (dashed curve) and 
  third (dotted curve) scenario halo masses were systematically under-
  and overestimated by 50\%.
}
\end{figure}

%
\begin{figure}
\resizebox{\hsize}{!}{\includegraphics[angle=-90]{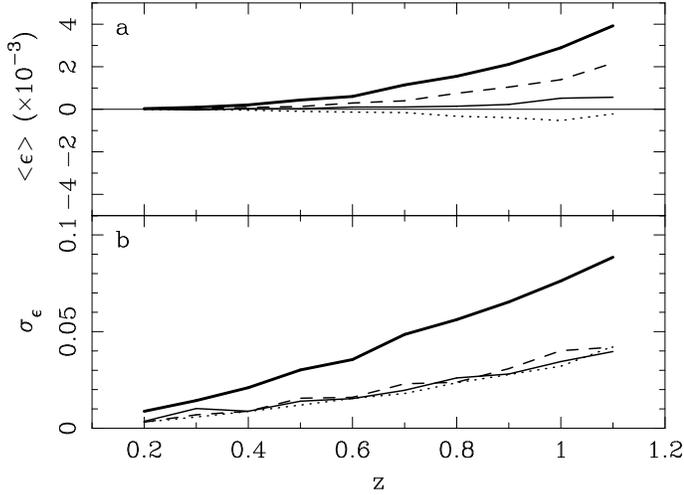}}
\caption{\label{fig:corrplot2}  Bias (panel a) and dispersion (panel b)
  in $\epsilon$ as a function of redshift.
  The no shift scenario is represented by
  the thin solid curves. Dashed and dotted curves correspond
  to the under- and overestimation scenarios, respectively.
  Bias and dispersion of $\Delta m_{\rm true}$,
  outlined by the thick solid curves, are also shown for comparison.
 To correct for gravitational lensing the formula
 $\Delta m_{\rm est}^{B}=B\Delta m_{\rm est}$ with $B$
 corresponding to the value outlined by the straight lines
 in Fig.~\ref{fig:scatter} was used. 
}
\end{figure}

%
\begin{figure}
\resizebox{\hsize}{!}{\includegraphics[angle=-90]{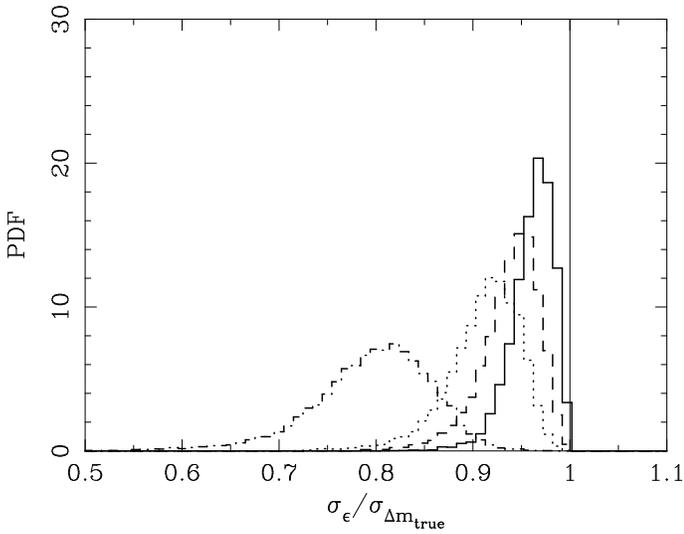}}
\caption{\label{fig:std2}  Probability distribution functions of 
  $\sigma_{\epsilon}/\sigma_{\Delta m_{\rm true}}$ for the no shift scenario 
  and different errors. 
  The distributions
  were obtained for
  simulated SNLS like data sets consisting of $250$ SNe~Ia corrected for gravitational 
  lensing using the formula $\Delta m_{\rm est}^{B}=B\Delta m_{\rm est}$. 
  The curves show results for different assumed errors:
  Solid curves correspond to $\sigma_{\rm err}$ given by Eq.~(\ref{eq:err}) and $\sigma_{\rm int}=0.13$ mag; 
  Dashed curves correspond to $\sigma_{\rm err}$ given by Eq.~(\ref{eq:err}) and $\sigma_{\rm int}=0.09$ mag; 
  Dotted curves correspond to $\sigma_{\rm err}=0$ and $\sigma_{\rm int}=0.09$ mag; 
  Dash-dotted curves correspond to $\sigma_{\rm err}=0$ and $\sigma_{\rm int}=0.05$ mag.
}
\end{figure}

\subsection{Judging the correction}
Correcting for gravitational lensing can of course only be justified
if there is a correlation between residuals and estimated
magnification. The presence or absence of a correlation after the
correction has been performed can be used to judge the success of the 
correction.
If the correction was successful, there should no longer be any
correlation. On the other hand, if the correction was not justified we
might have introduced a spurious correlation.

Although the slope differ for the different scenarios plotted in 
Fig.~\ref{fig:scatter}, the linear correlation
coefficient does not. The correlation coefficient, $r$, depends  on
the scatter, which is the same for all three scenarios.  
We have simulated
a large number of data sets and computed the correlation coefficient
before and after the correction.
Before the correction we consider the
correlation between $\Delta m_{\rm est}$ and Hubble diagram residuals.
After the correction the correlation between
$\Delta m_{\rm est}$ and $\epsilon=\Delta m_{\rm true}-\Delta m_{\rm est}$ 
is considered.  
For all three
scenarios, the initial correlation coefficient is 
$\langle r \rangle \simeq 0.27$.  The uncertainty in the
correlation coefficient decreases as $\sigma_r \simeq 1.3/\sqrt{N}$ as
the  number of SNe~Ia increases.  After applying the correction, 
we find $\langle r_{\rm corr}
\rangle$ to be approximately  $-0.02$, $0.09$, and $-0.10$ for the
no shift, under-, and overestimation scenario. For the under- and 
overestimation scenarios, where  the
magnifications are not correctly estimated, the correction hence
results in weak spurious correlations.  
Whether $r$ and $r_{\rm corr}$ are
significantly different or not depends on the sample size.  
For the no shift scenario, the difference would be significant (above
$3\sigma$) for $N \ga 200$.  
If the correction coefficient was taken into account,
$\Delta m_{\rm est}$ and
$\epsilon=\Delta m_{\rm true}-\Delta m_{\rm est}$ would be completely
uncorrelated since $B$ was fitted to the data, making this test 
inappropriate.

\section{Discussion and summary\label{sec:disc}}
Future SN~Ia data sets are anticipated to be large, which could justify
flux averaging as a method to overcome gravitational lensing bias.
However, such large homogeneous data sets will probably allow for a reduction 
of systematic uncertainties and the use of subsets of SNe~Ia -- or the discovery of 
new correlations with peak brightness -- may lead to 
large reductions of the intrinsic brightness dispersion.
Under such circumstances, 
gravitational lensing corrections may play an important role for 
precision cosmology.

In this paper we have investigated the possibility to perform corrections for gravitational 
lensing of individual SNe~Ia that reduces the scatter with negligible bias.
Since one of the largest uncertainties in the estimation of magnification
factors is the relation between luminosity and mass of galaxy haloes in the foreground,
 we have studied three different
scenarios: no shift in halo masses, underestimation by 50\% for all halo masses, and overestimation by
50\% for all halo masses. 
Our simulations show that for all three scenarios, the scatter due to gravitational lensing
can be reduced by roughly a factor 2 for a SNLS like data set. Also, the correction 
will be useful even if we do not know the luminosity-to-mass relation for the lensing galaxies to better than 50\%.
Any apprehension that gravitational lensing corrections
would lead to bias, thus appears to be unfounded. For simulated SNLS like data sets we found 
$|\langle \epsilon \rangle /\sigma_{\Delta m_{\rm true}}| \lesssim 0.25$.
The bias $\langle \epsilon \rangle$ is also expected to vary with redshift, but at a much smaller level
($|\langle \epsilon \rangle/\sigma_{\Delta m_{\rm true}}| \lesssim 4\times 10^{-2}$). 

As expected, the no shift scenario decreases the scatter more than the under- and overestimation scenarios.
However, a simple correction coefficient, $B$,  parameterizing the slope in the magnification-residual-diagram, 
can increase the performance of the corrections of the under- an overestimation scenarios
to the same level as the no shift scenario. Including the $B$-coefficient in the correction can 
hence eliminate the effect of bias in halo masses.

Correcting for gravitational lensing could reduce the total scatter in the SN~Ia magnitudes with $4\%$ for an intrinsic dispersion
of $\sigma_{\rm int}=0.13$ mag and errors similar to the ones obtained for SNLS \citep{ast06}. For $\sigma_{\rm int}=0.09$ mag, 
which might not be to unrealistic for future surveys, the total scatter could be reduced by $6\%$. Using Fisher matrix analysis
we find that the size of the error ellipses in the $(\Omega_{\rm M},w)$-plane can be reduced
by 4--8\%
for realistic measurement errors and realistic values of the intrinsic brightness scatter.

Gravitational magnification of SNe~Ia is rather independent of cosmological parameters. The 
cosmology dependence on the corrections should thus be rather small. However, when a 
more elaborate correction, such as Eq.~(\ref{eq:corrlog}), is employed which requires a fit to data
we have to be more careful. The residuals used depends on the cosmology and a proper 
correction must take this into account. One solution would be to iteratively compute the slope
in the magnification-residual-diagram and estimate the cosmological parameters.
In passing, we note that the gravitational lensing magnification
 we consider here affects all points of the SN~Ia light-curve by the same amount, making it 
possible to apply the corrections after the light-curve fit. 

Corrections for gravitational lensing could also be important for other distance indicators.
Gravitational waves emitted by chirping binary systems could -- if detected -- be used to obtain 
very accurate luminosity distances \citep{sch86}. Furthermore, if the redshift of the binary could
be measured from an optical counterpart, these binary system could be standard sirens, the 
gravitational wave analogs of standard candles. These standard sirens are unaffected by most 
systematic uncertainties which plague standard candles and they might  
provide distances with relative accuracy  $\lesssim 1\%$. Standard sirens and standard 
candles both suffer extra dispersion due to gravitational lensing. For standard sirens this 
effect is much more important than for standard candles, since the other uncertainties are 
so small. Gravitational lensing will thus degrade the power of chirping binary systems as
distance indicators \citep{hol05b}. The potential improvement of correcting standard sirens
for lensing was investigated in  \citet{jon07}. Corrections could restore some 
of their power. The results found here that gravitational lensing corrections can be unbiased 
could thus be of importance also for future precision gravitational wave cosmology.

\begin{acknowledgements}
The Dark Cosmology Centre is funded by the 
Danish National Research Foundation. 
EM and JS acknowledges financial support from the Swedish
Research Council and from the Anna-Greta and Holger Crafoord fund.
JS is a Royal Swedish Academy of Sciences Research Fellow supported 
by a grant from the Knut and Alice Wallenberg Foundation. 

\end{acknowledgements}



\begin{thebibliography}{}
%

\bibitem[Astier et al.(2006)]{ast06}
Astier, P., Guy, J., Regnault, N., et al. 2006, A\&A, 447, 31

\bibitem[Cooray et al.(2006)]{cor06}
Cooray, A., Huterer, D., \& Holz D. 2006, Phys. Rev. Lett., 96, 021301

\bibitem[Dahl\'en et al.(2005)]{dah05}
Dahl\'en, T., Mobasher, B., Somerville, R.~S., et al. 2005, ApJ, 631, 126 

\bibitem[Dalal et al.(2003)]{dal03}
Dalal, N., Holz, D.~E., Chen, X., Frieman, J.~A.
2003, ApJ, 585, L11

\bibitem[Frieman et al.(2008)]{fri07}
Frieman, J.~A., Basset, B., Becker, A., et al. 2008, 
AJ, 135, 338    

\bibitem[Gunnarsson(2004)]{gun04}
Gunnarsson, C. 2004, JCAP, 03, 002

\bibitem[Gunnarsson et al.(2006)]{gun06}
Gunnarsson, C., Dahl\'en, T., Goobar, A., J\"onsson, J., \& M\"ortsell, E. 2006, ApJ, 640, 471

\bibitem[Holz \& Hughes(2005)]{hol05b} Holz, D. \& Hughes, S. 2005, ApJ, 629, 15 

\bibitem[Holz \& Linder(2005)]{hol05}
Holz, D.~E. \& Linder, E.~V. 2005, ApJ, 631, 678

\bibitem[J\"onsson et al.(2007)]{jon07}
J\"onsson, J., Goobar, A., \& M\"ortsell, E. 2007, ApJ, 658, 52

\bibitem[J\"onsson et al.(2008)]{jon08}
J\"onsson, J., Kronborg, T., M\"ortsell, E., \& Sollerman, J. 2008, 
A\&A, 487, 467

\bibitem[Leibundgut(2008)]{lei08} 
Leibundgut, B. 2008, General Relativity and Gravitation, 40, 221

\bibitem[Miknaitis et al.(2007)]{mik07}
Miknaitis, G, Pignata, G., Rest, A., et al. 2007, ApJ, 666, 674 

\bibitem[Navarro et al.(1997)]{nav97}
Navarro, J.~F., Frenk, C.~S., \& White, S.~D.~M. 1997, ApJ, 490, 493

\bibitem[Sarkar et al.(2008)]{sar07}
Sarkar, D., Amblard, A., Holz, D., \& Cooray, A. 2008, ApJ, 678, 1

\bibitem[Schutz(1986)]{sch86} Shutz, B. 1986, Nature, 323, 310

\bibitem[Wang(2000)]{wan00}
Wang, Y. 2000, ApJ, 536, 531

\bibitem[Wang \& Mukherjee(2004)]{wan04}
Wang, Y.  \& Mukherjee, P. 2004, ApJ, 606, 654

\bibitem[Wood-Vasey et al.(2007)]{woo07}
Wood-Vasey, W.~M., Miknaitis, G., Stubbs, C.~W., et al. 2007, ApJ, 666, 694

\end{thebibliography}
\end{document}